\documentclass[conference]{IEEEtran}
\usepackage{caption}
\usepackage[sorting=none]{biblatex}                 
\usepackage{csquotes}                               
\usepackage[hidelinks]{hyperref}                    
\usepackage{xurl}
\hypersetup{breaklinks=true}

\ifCLASSINFOpdf
\else
\fi
\usepackage{multirow}
\usepackage{graphicx}
\usepackage{listings}
\usepackage[dvipsnames]{xcolor}

\usepackage[cmex10]{amsmath}

\usepackage{verbatim}

\usepackage{authblk}

\title{Harnessing Large Language Models for Software Vulnerability Detection: A Comprehensive Benchmarking Study}

\author[1]{Karl Tamberg}
\author[1, 2]{Hayretdin Bahsi}

\affil[1] {School of Information Technologies\\
Tallinn University of Technology}
\affil[2] {School of Informatics, Computing, and Cyber Systems\\
Northern Arizona University}

\affil[ ]{\textit{katamb@taltech.ee, hayretdin.bahsi@taltech.ee}}

\begin{document}

\IEEEoverridecommandlockouts

\maketitle

\begin{abstract}
Despite various approaches being employed to detect vulnerabilities, the number of reported vulnerabilities shows an upward trend over the years. This suggests the problems are not caught before the code is released, which could be caused by many factors, like lack of awareness, limited efficacy of the existing vulnerability detection tools or the tools not being user-friendly. To help combat some issues with traditional vulnerability detection tools, we propose using large language models (LLMs) to assist in finding vulnerabilities in source code. LLMs have shown a remarkable ability to understand and generate code, underlining their potential in code-related tasks. The aim is to test multiple state-of-the-art LLMs and identify the best prompting strategies, allowing extraction of the best value from the LLMs. We provide an overview of the strengths and weaknesses of the LLM-based approach and compare the results to those of traditional static analysis tools. We find that LLMs can pinpoint many more issues than traditional static analysis tools, outperforming traditional tools in terms of recall and F1 scores. The results should benefit software developers and security analysts responsible for ensuring that the code is free of vulnerabilities.
\end{abstract}

\section{Introduction}
As software is integrated into many business processes, a substantial amount of code is written every day. High-level languages and frameworks take some of the responsibility off the developers' shoulders by introducing various functional and security components into the software. For instance, higher-level languages add abstraction and take care of memory management behind the scenes, improving the security posture. However, this abstraction does not address all vulnerabilities. 

The records hosted in the Common Vulnerabilities and Exposures (CVE) database show an increasing trend of reported vulnerabilities over the years \cite{cve-about}. In the past five years alone, nearly 110,000 CVEs have been reported \cite{cve-about}. Even though this increase may indicate a more thorough security analysis conducted by the security communities after software releases, such high numbers threaten the security landscape and underline the need for developing better vulnerability detection tools and techniques that are applicable before software deployments. 

Many efforts have been made in this field, including automating some software development and deployment processes. Automatic vulnerability detection can be integrated into both the development tooling and code review processes. Most tools work as black boxes, meaning there is no need for the user to understand how the tool works; the user just provides the code and receives the results. However, the techniques used to detect vulnerabilities or bad coding practices can be ineffective \cite{sast-ineffective, static-anal-cwe-grouping}. Dynamic analysis techniques suffer from performance problems and have a high tendency to miss problems \cite{sast-ineffective}. Some existing static analysis techniques suffer from high false positive rates and others from only being able to detect some specific vulnerabilities \cite{sast-ineffective}.

The code generation capabilities of large language models (LLMs) have already been recognised, and a commercial tool, Codex, has been developed \cite{openai-codex}. The extraordinary code-understanding capability of LLMs makes them a favourable solution for the detection of vulnerabilities in software code. As an additional advantage, follow-up questions can be directed at the LLMs. Thus, their capabilities can be extended to suggest and implement fixes.

Prompting strategies allow models with larger parameter sizes to achieve very high performance in handling novel tasks without fine-tuning \cite{human-language}. While multiple studies have tested the vulnerability detection capabilities of LLMs, they do not conduct a comprehensive benchmarking of prompting techniques \cite{understanding-llm-effectiveness, vul-detect-prompt-enhanced, noever2023-llm-fix, llms-emerge-vuln-detect}. More specifically, they do not include some of the state-of-the-art prompting techniques (e.g., tree of thoughts (ToT) \cite{tot-prompt}, self-consistency \cite{self-consistency}) and recent large language models (e.g., Claude 3 Opus model). Although the cost of API utilization is a significant factor for widely used commercial LLM models, they do not provide a cost analysis in their benchmarks. Their comparison with the baseline static analysis tools is not rigorous. They often perform a simpler classification task, such as evaluating whether a vulnerability exists in a code snippet or not \cite{understanding-llm-effectiveness, vul-detect-prompt-enhanced}, requiring extra effort to detect the exact issue.

In this paper, we conduct a comprehensive benchmarking study of the state-of-the-art and off-the-shelf large language models (LLMs), treating them as black-box static code analysers. We formulate a multi-class classification task in which the prompts ask LLMs about the vulnerability type that the given code snippet may have. The benchmarking includes various prompting techniques and their combinations to identify the best-performing ones. We compare the LLM results with static code analysis tools and provide a pros and cons analysis between these two alternatives. Based on our findings, we present some practical suggestions for security practitioners about more effective utilisation of LLMs in vulnerability detection tasks.

The research questions we aim to answer are: (1) What prompting approach is most successful with the LLMs to detect vulnerabilities? (2) What advantages and disadvantages do LLMs have over existing static analysis tools? (3) How would the use of off-the-shelf LLMs be able to contribute to vulnerability detection in source code?

We use a synthetic dataset called Juliet Java 1.3 \cite{juliet-java-13} and apply a rigorous pre-processing stage to eliminate textual clues about the vulnerability type and presence. As a baseline comparison, we use two state-of-the-art static code analysis tools, CodeQL \cite{github-codeql} and SpotBugs \cite{spotbugs}. SpotBugs has shown good results in comparison to other traditional static analysis tools on the Juliet dataset \cite{pmd-spotbugs-juliet, juliet-java-sat-eval} and CodeQL has been previously utilised as a comparison point for testing LLM capabilities \cite{understanding-llm-effectiveness}. For the initial experiments, the GPT-4 turbo model from OpenAI is utilised, as it has a relatively large context window, is relatively cheap to use and is preferred by the community \cite{chatbot-leaderboard}. Then, we run the best-performing strategies on two more models, GPT-4 and Claude 3 Opus, which are voted to be among the best ones available at the time of running the experiments \cite{chatbot-leaderboard}. 

The main contribution of this paper is outlined as follows:
\begin{itemize}

\item  We include prompting strategies not previously considered for vulnerability detection tasks, such as the tree of thoughts (ToT) \cite{tot-prompt} and self-consistency \cite{self-consistency}. We demonstrate the results of experiments with various combinations of prompting techniques. 

\item We propose a novel prompt approach containing detailed security review instructions that outperform other previously proposed strategies in most settings. 

\item The results are compared with two traditional static analysis tools, CodeQL \cite{github-codeql} and SpotBugs \cite{spotbugs}. The comparison includes the best-performing configurations of these tools. Similar studies have either not used traditional tools as a comparison point \cite{java-gpt-eval, cot-vsp-llms} or have not elaborated on the exact configurations and optimisations of the tools used \cite{understanding-llm-effectiveness}.   

\item We provide detailed costs per prompting strategy and discuss the cost as a factor in choosing LLMs and prompting strategies. Cost has not been explored in detail by prior studies in the same domain.

\item We prepare prompts for LLMs to induce a multi-class classification task that identifies the vulnerability type based on the CWE category. Some of the similar studies induce binary classification tasks, requiring more prompts for CWE identification \cite{understanding-llm-effectiveness, vul-detect-prompt-enhanced, llms-emerge-vuln-detect}.  

\item We include the Claude 3 Opus model from Anthropic in our benchmarking. To the best of our knowledge, we are the first to include this model in a vulnerability detection benchmark study.

\item We discuss the strengths and weaknesses of LLM-based vulnerability detection in comparison to traditional static analysis tools. 

\end{itemize}

The content of the paper is given as follows: Section \ref{_B_and_RW_} gives background information and reviews the literature. The methods applied in this study are detailed in Section \ref{_Methods_}. Section \ref{_Results_} presents the benchmarking results. The main findings and limitations of this study are elaborated in Section \ref{_Discussion_}. Section \ref{_Conclusion_} concludes the paper.

\section{Background and Related Work}\label{_B_and_RW_}
\subsection{Background}
The difficulty of writing code without any flaws, which could introduce vulnerabilities, has been recognised and many different approaches to help developers find issues in their code have been proposed. These approaches include manual code reviews, penetration testing, using static/dynamic/hybrid code analysers, machine learning and deep learning strategies.

Manual security code review is a form of static code analysis, where humans check code for security vulnerabilities. This is very time-consuming and expects the reviewer to know what vulnerabilities to look for \cite{auto-code-review}. While manual reviews can be effective, relying on manual reviews alone has been shown to not be reliable \cite{effectiveness-manual-code-review}.
 
Manual penetration testing consists of manually testing the application during runtime to find or confirm security issues. This approach is very time-consuming, relies on the tester to be familiar with all the application flows and is likely to miss some vulnerabilities \cite{pen-testing-based-on-threat-model}.

Automated approaches allow the discovery and flag of potential vulnerabilities without manual effort. Automated static code analysis applies pre-defined rules or algorithms on the source code, deriving a list of potential vulnerabilities \cite{mitigating-program-sec-vuln}. In theory, most vulnerabilities could be detected with static analysis techniques \cite{mitigating-program-sec-vuln}. In practice, static code analysis tools are limited by the vulnerability types they can detect \cite{sast-ineffective}. The advantages of static analysis include not having to execute the code, meaning there are no issues related to the reachability of vulnerabilities (i.e., reachability means all of the code in the codebase can be analysed) \cite{mitigating-program-sec-vuln}. Static analysis allows for a quick turnaround for fixes, as it can point to the exact location in the code where the vulnerability lies \cite{akcurastatic}. What is more, the tooling is often simple and relatively fast to use, which means the code analysis can be done even on local machines. 

Dynamic code analysis is the analysis of code behaviour in runtime \cite{taxonomy-scat}. This approach consists of running the program in specific circumstances, during which the behaviour of the program is monitored. Dynamic analysis approaches can struggle with reachability, meaning not all of the code can be analysed \cite{mitigating-program-sec-vuln}. Dynamic code analysis is expected to have fewer false positives than static analysis, but more false negatives \cite{akcurastatic}. If dynamic analysis uncovers an issue, it can be hard to trace the issue back to specific parts of the code \cite{akcurastatic}.

Hybrid code analysis is a mixture of static and dynamic approaches, allowing them to combine their results for better outcomes. For example, a hybrid code analysis approach could use the warnings from the static analysis tools as input and run dynamic analysis to either verify or discard these warnings \cite{mitigating-program-sec-vuln}. 
The previous example's flaw is that it will not uncover additional vulnerabilities compared to static analysis alone, but it can filter out some false positives.

Machine learning approaches have been tested for vulnerability detection tasks, however, the proposed approaches have many drawbacks. Some require domain experts to be part of the feature extraction process, which can be time-consuming, error-prone and task-specific \cite{ml-dl-survey}. Some have not shown good results in vulnerability detection tasks \cite{soft-vuln-anal-ml-survey} and others can only be used for detecting small sets of vulnerability types \cite{lit-survey-dl-vuln-anal}.

To help combat the problems of ML-based vulnerability detection approaches, multiple deep learning (DL) approaches have been proposed. The layered structure of DL models is claimed to be better at capturing complex patterns in source code \cite{ml-dl-survey}. What is more, the DL approaches allow the feature extraction processes to be automated, requiring less manual labour-intensive tasks \cite{ml-dl-survey}. The main issues with current deep learning vulnerability detection systems are the focus on a single programming language and the use of API function calls to locate vulnerabilities \cite{lit-survey-dl-vuln-anal}. Furthermore, the datasets used in DL vulnerability prediction studies are often too simple and do not translate to real-world use cases, suffering from imbalanced data and failing to address code semantics \cite{dl-vul-detection-whadup}.

\subsection{Related work}
LLMs have shown good results when tested for vulnerability detection purposes. 
When GPT-4 is compared to CodeQL on OWASP and Juliet Java datasets, they show similar results, with GPT-4 performing better on the OWASP dataset and CodeQL performing better on the Juliet Java dataset \cite{understanding-llm-effectiveness}. For prompting, four approaches are discussed in detail: basic prompt, CWE-specific prompt, dataflow analysis prompt and dataflow analysis prompt with self-reflection. The dataflow analysis prompt with self-reflection shows the best results for binary vulnerability detection \cite{understanding-llm-effectiveness}.
On Java and C/C++ datasets, GPT-4 is shown to outperform existing deep learning tools even with basic prompts \cite{vul-detect-prompt-enhanced}. Role-based prompting improves results, but GPT-4 shows bias towards prompt wording, favouring responses aligning with the prompt \cite{vul-detect-prompt-enhanced}. More complex strategies involving API call sequences and dataflow descriptions are explored, with the order of prompt components impacting effectiveness and the prompt containing the API call sequence showing the best results \cite{vul-detect-prompt-enhanced}.
The GPT-4 reportedly outperforms static analysis tools like Snyk and Fortify when tested on real-world scientific repositories \cite{noever2023-llm-fix}. It is discovered that asking the LLM to provide a fix for the detected vulnerability improves the detection capabilities, most likely due to forcing the model to explain and justify its response \cite{noever2023-llm-fix}.
Few-shot prompting has also shown promise for vulnerability detection tasks when used with GPT-4 \cite{llms-emerge-vuln-detect}. 
The GPT-4 vulnerability detection capabilities for OWASP's top 10 vulnerabilities are tested on a purpose-made PHP dataset \cite{chatbot-to-replace-sat}. The results are compared to those of 11 different static analysis tools and the GPT-4 outperforms all in terms of true positive rate while suffering from high false positive results.
Few-shot CoT approaches have been proposed for different vulnerability classification tasks \cite{cot-vsp-llms}. The LLM is asked to only focus on relevant parts of the code and the approach is dubbed as vulnerability semantics guided prompting (VSP), outperforming other tested strategies \cite{cot-vsp-llms}.

Other attempts have shown less success in utilising LLMs as black-box static code analysers. 
The authors of Codex, the LLM that also powers Github Copilot, also consider applying the Codex model for vulnerability discovery \cite{evaluating-codex}. While no benchmarking results or techniques are discussed, it is claimed their testing does not reveal any cases where Codex outperforms static analysis tools \cite{evaluating-codex}. However, more capable models are recognised as potentially performing better in identifying vulnerabilities and the need for further research in the area is emphasized \cite{evaluating-codex}.
GPT-3 and GPT-3.5 on a real-world Java dataset containing 120 samples fail to outperform a dummy classifier \cite{java-gpt-eval}. However, it must be noted the authors test only basic zero-shot prompting techniques and acknowledge that newer models and better prompting techniques could improve the results \cite{java-gpt-eval}.
When the focus is on the quality of the output from LLMs, it is found that LLMs can struggle to provide correct, understandable, concise, consistent and compliant responses \cite{sec-code-review-llm-responses}. A prompt asking the LLM to find vulnerabilities and classify them as one of the 10 high-level research view vulnerabilities (from CWE-1000) is shown to perform the best \cite{sec-code-review-llm-responses}. Asking about high-level vulnerabilities could be one reason for the vagueness in the responses.

Most prior work on the subject does not approach the task systematically. The challenge is often formulated as a binary classification challenge \cite{understanding-llm-effectiveness, vul-detect-prompt-enhanced, llms-emerge-vuln-detect}. The studies considering multi-class classification do not discuss using any matching strategies to be able to consider multiple CWE categories correct for some vulnerabilities \cite{chatbot-to-replace-sat}. The more elaborate prompting strategies from the literature like ToT \cite{tot-prompt} or self-consistency \cite{self-consistency} are not tested and while the CoT prompting is tested by some, these CoT prompts contain few broad steps instead of detailed instructions \cite{vul-detect-prompt-enhanced}. The attention is often disproportionately directed towards true positives, while there is limited focus on the false positives \cite{noever2023-llm-fix, chatbot-to-replace-sat}.

\section{Methods}\label{_Methods_}
\subsection{LLMs and prompting strategies}
To conduct static application security testing with LLMs, we used three models, GPT-4 and GPT-4 turbo models from OpenAI\footnote{\url{https://openai.com/}} and Claude 3 Opus model from Anthropic\footnote{\url{https://www.anthropic.com/}}. The exact versions for the LLMs and other relevant tools are given in ~\nameref{appendix:versions}. For most experiments, the temperature parameter of the LLM is set to zero. The temperature parameter is used to control the randomness of the output and using a zero value makes the output as deterministic as possible \cite{openai-api-doc-chat}. This means anyone running the same experiments with the same settings should get very similar results to us.

LLMs require a description of the task they are expected to perform and these descriptions are called prompts. It has induced a separate field of study to find the best prompting approaches for different types of tasks. Zero-shot prompting stands for an approach where the machine learning model might not be trained for such a task and the prompt does not include examples \cite{zero-shot-learning-eval}. Few-shot prompting stands for an approach where the LLM is offered a few examples in the prompt, often together with a task description \cite{logan2021cutting}. Chain of thought (CoT) prompting has been proven to work well when asking LLMs to solve complex problems which are reasonable to tackle in multiple steps \cite{chain-of-thought}. Further development of the CoT prompting called the tree of thoughts (ToT) prompting has been developed, where models evaluate various reasoning paths and self-assess their choices \cite{tot-prompt}. Self-consistency involves sampling diverse outputs from a language model and then selecting the most consistent answer from that set \cite{self-consistency}.

\subsection{Dataset}
The full Java Juliet 1.3 dataset contains vulnerabilities from 112 different CWEs. Among others, the dataset covers categories like CWE-546, which refers to suspicious comments. To focus on high-severity issues, only vulnerabilities from the MITRE top 25 \cite{mitre-top-25} are chosen, similarly to previous research \cite{understanding-llm-effectiveness}. It must be noted the CWE list is very detailed and many weaknesses in that list are very similar or closely related. To help understand how different CWEs are related, MITRE provides multiple views, which show the relationships between CWEs. The "Research Concepts" (CWE-1000) view provided by MITRE displays the relationships in a hierarchy, where every CWE can have one parent and multiple children \cite{mitre-cwe-1000}. As the Java Juliet 1.3 dataset only contains four CWEs from the 2023 MITRE top 25 list \cite{mitre-top-25}, we also include the subcategories of the MITRE top 25 vulnerabilities. For example, the CWE-22 (Path Traversal) is in the MITRE top 25 list but not in the Juliet dataset \cite{mitre-top-25}. However, the Juliet dataset contains the subcategories of CWE-22, namely CWE-23 (Relative Path Traversal) and CWE-36 (Absolute Path Traversal), which we include. Using this strategy allows us to extend our dataset from the four exact matches to 17 distinct CWEs.

The default distribution of the Java Juliet 1.3 dataset is using Ant\footnote{\url{https://ant.apache.org/}} as the build tool, which we upgrade to Gradle\footnote{\url{https://gradle.org/}}. This update facilitates the easier integration of traditional static analysis tools. The code reformatting tools included in IntelliJ\footnote{\url{https://www.jetbrains.com/idea/}} are used to reformat the files, which helps to ensure all files use a similar format. 

Prior research has shown leaking some relevant keywords in the code, like variables named "secure", could influence the output of the LLMs \cite{LLbezpeky}. To avoid introducing this bias, these types of hints are removed from the dataset. The original dataset contains comments explaining the vulnerabilities, so all comments are removed. As the file and class name both include the vulnerability identifier, all the files and classes are re-named. All the function and variable names containing any hints (like "good" or "bad") are also changed. The package name is not changed to simplify the analysis of the results. However, the package name is always overwritten when working with LLMs, before the file is sent to the LLM for analysis. To give as much context to the LLMs for vulnerability detection as possible, we test the detection capabilities on the file level. By default, the Juliet dataset is configured for function or file-level vulnerability detection. Similarly to previous research, the non-standard test cases spanning multiple files or only containing vulnerable examples are removed \cite{understanding-llm-effectiveness}. The remaining samples are split into two: a good and a bad file. This way, all the context needed to detect a vulnerability is contained inside a single file.

After removing the test cases spanning multiple files and splitting the remaining files into two, we are left with 15,174 files, half of them secure and the other half vulnerable. Due to the high cost of running LLMs, we are unable to experiment with all of the files. Thus, as the last step, a random subset of the files is selected. We select 17 vulnerable and 17 non-vulnerable files for each of the 17 CWE categories at random. Altogether $(17 + 17) \times 17 = 578$ files are chosen. 

The full pre-processed dataset is available in GitHub\footnote{https://github.com/katamb/juliet-top-25}. The custom scripts used for the pre-processing are available in GitHub\footnote{https://github.com/katamb/thesis-scripts} under the "dataset-normalization" package.

\subsection{Result evaluation}
We formulate vulnerability detection as a multi-class classification task. The analysis is expected to report not only whether the file is vulnerable, but also to correctly detect the CWE identifier of the vulnerability. The Juliet dataset is labelled, which allows us to classify the results as true positive, false positive, true negative or false negative. If the expected vulnerability is detected, then the classification is positive.

Using those measures, we can compare the performance of different tools by calculating accuracy, precision, recall and F1 (harmonic mean of recall and precision) scores. The formulas to calculate these values are as follows \cite{prec-rec-f1-calculations}:
\begin{minipage}{0.5\textwidth}
\begin{align*}
    Accuracy &= \frac{TP + TN}{TP + FP + TN + FN}\\
    Precision &= \frac{TP}{TP + FP}\\
    Recall &= \frac{TP}{TP + FN}\\
    F1 &= 2 \times \frac{precision \times recall}{precision + recall}.
\end{align*}
\end{minipage}
\vspace{1.5mm}

The authors of the Juliet dataset acknowledge the code provided in the dataset might include other unrelated vulnerabilities \cite{juliet-12-user-guide}. Thus, for vulnerable files, the result is considered true positive only when the targeted vulnerability is found in the file. If no vulnerabilities are found or if the found vulnerabilities do not include the targeted vulnerability, the result is classified as a false negative. For non-vulnerable files, the result is considered true negative if the targeted vulnerability is not discovered in the file. If the targeted vulnerability is found, the result is classified as a false positive.

Many CWEs point to very similar flaws. The "Research Concepts" (CWE-1000) view provided by MITRE gives a good overview of the relationships between vulnerabilities \cite{mitre-cwe-1000}. For example, the CWE-36 (Absolute Path Traversal) is present in our dataset. The parent and the children of CWE-36 all point to different variations of absolute path traversal weakness. To fairly assess the results, we employ a strategy similar to what has been used for evaluating traditional static analysis tools previously \cite{static-anal-cwe-grouping, static-anal-cwe-grouping-2}. This strategy allows the parent CWE and the child CWEs to also be considered to be a correct classification based on the MITRE "Research Concepts" view. The only exception we make is related to the highest level CWEs in the "Research Concepts" view, which are called pillars. If the parent of the CWE is of a type pillar, we do not count the parent as the correct classification, as the pillar descriptions can be very broad. For example, the parent of CWE-476 (NULL Pointer Dereference) is CWE-710 (Improper Adherence to Coding Standards), which contains many different weaknesses. Interestingly, even though this strategy has been employed in static code analyser benchmarks \cite{static-anal-cwe-grouping, static-anal-cwe-grouping-2}, it has not been previously utilised in LLM benchmarking studies \cite{understanding-llm-effectiveness, vul-detect-prompt-enhanced, noever2023-llm-fix, llms-emerge-vuln-detect}. This is most likely related to either a lack of awareness or the complexity it adds to the result evaluation process.

\section{Results}\label{_Results_}
\subsection{Traditional static analysis tools}
The traditional static analysis tools chosen are CodeQL and SpotBugs, as both have been shown to perform well on synthetic datasets \cite{sast-comparison}. Like most static code analysis tools, both CodeQL and SpotBugs need the dataset to be compileable to run the scan. The results from these tools serve as a comparison point to better evaluate the results produced by LLMs. CodeQL has been utilised by previous similar studies as the comparison point as well \cite{understanding-llm-effectiveness}.

CodeQL is written and maintained by GitHub and the community, with the queries being open-source \cite{github-codeql-about}. Out of the box, CodeQL provides three different configurations for Java code analysis: the default configuration, the extended security configuration, and the extended security and quality configuration \cite{github-codeql-about}. CodeQL provides mappings for CWE-IDs, which significantly simplifies analyzing the results \cite{github-codeql-cwe-mapping}. All three different configurations are tested, with the results displayed in Table \ref{tab:sast_results}. The default configuration is denoted as CodeQL-d, the extended security configuration is denoted as CodeQL-es and the extended security and quality configuration is denoted as CodeQL-esq. The default configuration seems to be configured to produce as good precision as possible. However, the other configurations perform better in terms of accuracy, recall and F1 values. The extended security and quality configuration produces the best results, achieving an F1 score of 0.61 (see Table \ref{tab:sast_results}). Unfortunately, the previous study benchmarking LLM vulnerability detection capabilities against CodeQL does not disclose the exact tested or used configurations \cite{understanding-llm-effectiveness}. 

\begin{table*}[htbp]
    \renewcommand{\arraystretch}{1.1}
    \caption{\it{Static analyser results}}
    \label{tab:sast_results}
    \centering
    \begin{tabular}{|c||c|c|c|c||c|c|c|c|}
    \hline
        & \textbf{TP} & \textbf{FP} & \textbf{TN} & \textbf{FN} & \textbf{Accuracy} & \textbf{Precision} & \textbf{Recall} & \textbf{F1} \\
        \hline
        CodeQL-d & 76 & 5 & 284 & 213 & 0.623 & \textbf{0.938} & 0.263 & 0.411 \\
        \hline
        CodeQL-es & 127 & 20 & 269 & 162 & 0.685 & 0.864 & 0.439 & 0.583 \\
        \hline
        CodeQL-esq & 137 & 23 & 266 & 152 & 0.697 & 0.856 & 0.474 & 0.61 \\
        \hline
        SpotBugs-d & 39 & 20 & 269 & 250 & 0.533 & 0.661 & 0.135 & 0.224 \\
        \hline
        SpotBugs-fsb & 152 & 23 & 266 & 137 & \textbf{0.723} & 0.869 & \textbf{0.526} & \textbf{0.655} \\
        \hline
    \end{tabular}
\end{table*}

SpotBugs is a tool for finding bugs in Java code \cite{spotbugs-intro}. SpotBugs supports plugins, from which the Find Security Bugs plugin is used, similarly to prior benchmarks \cite{sast-comparison}. The results of running the analysis are summarised in Table \ref{tab:sast_results}. The default configuration is denoted as SpotBugs-d, and the configuration with the Find Security Bugs plugin is denoted as SpotBugs-fsb. The documentation only provides some CWE mappings for the vulnerabilities detected by the Find Security Bugs plugin; the rest require manually mapping the results to CWE identifiers. The use of Find Security Bugs significantly improves the vulnerability detection capabilities of the SpotBugs tool. The SpotBugs-fsb configuration performs better than the default configuration for accuracy, precision, recall and F1 scores as shown in Table \ref{tab:sast_results}. It also performs better than the best CodeQL approach, showing higher scores in all measurements. 

\subsection{LLMs as static analysis tools}
For the experiments, we use the GPT-4 turbo model with the temperature set to 0 unless explicitly stated otherwise. There are two reasons for opting to use the GPT-4 turbo model for the majority of the tests. Firstly, the GPT-4 turbo model costs per token are significantly cheaper than those of the GPT-4 or Claude 3 Opus. Secondly, the GPT-4 turbo has a larger context window. The temperature parameter is used to control the randomness of the output and using a zero value makes the output as deterministic as possible \cite{openai-api-doc-chat}. This means anyone running the same experiments with the same settings should get very similar results to us.

It is important to note that the LLMs are accessed through an API and are used as a black box. OpenAI and Anthropic APIs are used in conjunction with the LangChain\footnote{\url{https://github.com/langchain-ai/langchain}} Python library to conduct the experiments. Both OpenAI and Anthropic use a token-based pricing structure, which means they request money for every input and output token. For the models used, output tokens cost between two to five times more than input tokens. This provides an incentive to give as much context as possible to the model as input. 
The result tables contain cost and time columns, which are aggregated values over the whole dataset (578 files). The cost is in dollars and excludes VAT. To save on the costs, we remove the indentation from the Java files before sending them to the LLM. The time is marked in hours and it must be noted the time it takes to run the same prompt through the same LLM seems to vary notably. 
Thus, the time noted here is not a reliable measure but more of an indication of how much time the analysis took in our settings and gives insights about a rough comparison between different models and prompting alternatives. 

\subsubsection{Response re-evaluation strategies}

To first establish a baseline, a basic prompt, which we denote as $p_b$, is compiled using the best practices suggested by OpenAI for prompt engineering \cite{openai-prompt-eng}. The LLM is asked to adopt the persona of a security researcher, and the instructions are clear on what is expected and what the output should be.

There have been suggestions to ask LLMs to re-evaluate their responses to achieve better results, and different approaches for this have been proposed \cite{understanding-llm-effectiveness, openai-prompt-eng, recursive-criticism-and-improvement, self-refinement-prompting}. Four different approaches are tested, where the output of the basic prompt is given back to the LLM and the LLM is asked to re-evaluate and improve its previous response. All the approaches are quite similar, but they focus on slightly different aspects. The first approach is called recursive criticism and improvement (RCI), which we denote as $p_{b-rci}$ \cite{recursive-criticism-and-improvement}. RCI builds on the initial prompt and answer from the LLM, asking the LLM to find problems with its initial answer and then improve it based on the problems. The second improvement tactic is called self-refinement, denoted as $p_{b-sr}$ \cite{self-refinement-prompting}. Self-refinement asks to provide overall feedback (instead of just focusing on problems) on the previous answer and to improve the initial response. For the third approach, we asked the LLM to provide feedback and response in one go and called it short self-refinement ($p_{b-ssr}$). The short self-refinement has the advantage of being faster and cheaper, as the LLM is invoked only twice and fewer tokens are used. This approach is the most similar to what has been utilised by previous research on the topic \cite{understanding-llm-effectiveness}. As can be seen in Table \ref{tab:llm_results}, the RCI strategy performed the best. To test if we could achieve similar results with cheaper costs, lastly, a strategy we call short RCI ($p_{b-srci}$) is tried. The idea behind short RCI is to try and find out how the criticism and improvements in one step perform in comparison to the original. To test if we could achieve similar results with cheaper costs, lastly, we combine the criticism and improvement steps into one step in an approach we call short RCI ($p_{b-srci}$). As seen in Table \ref{tab:llm_results}, the short RCI tactic ($p_{b-srci}$) shows second-best results. While the short RCI tactic does provide significant time and cost savings in comparison to the full RCI, it performs more poorly in all other aspects. We hypothesize the difference in results is related to the RCI prompt, allowing the LLM to generate relevant clues and build corrections for those clues. Asking the LLM to do both in the same step will give the LLM less input for the final verdict, which seems to affect the results negatively.

\begin{table*}[htbp]
    \renewcommand{\arraystretch}{1.1}
    \caption{\it{LLM results}}
    \label{tab:llm_results}
    \centering
    \begin{tabular}{|c||c|c|c|c||c|c|c|c||c|c|}
        \hline
        \textbf{Strategy} & \textbf{TP} & \textbf{FP} & \textbf{TN} & \textbf{FN} & \textbf{Accuracy} & \textbf{Precision} & \textbf{Recall} & \textbf{F1} & \textbf{Cost} & \textbf{Time} \\
        \hline
        $p_b$ & 134 & 131 & 158 & 155 & 0.505 & 0.506 & 0.464 & 0.484 & 4.38\$ & 0.9h \\
        \hline
        $p_{b-rci}$ & 140 & 49 & 240 & 149 & 0.657 & 0.741 & 0.484 & 0.586 & 17.47\$ & 4.2h \\
        \hline
        $p_{b-sr}$ & 135 & 91 & 198 & 154 & 0.576 & 0.597 & 0.467 & 0.524 & 22\$ & 6.8h \\
        \hline
        $p_{b-ssr}$ & 119 & 70 & 219 & 170 & 0.585 & 0.63 & 0.412 & 0.498 & 9.72\$ & 1.9h \\
        \hline
        $p_{b-srci}$ & 133 & 80 & 209 & 156 & 0.592 & 0.624 & 0.46 & 0.53 & 11.07\$ & 2.6h \\
        \hline
        $p_{b-sc}$ & 133 & 133 & 156 & 156 & 0.5 & 0.5 & 0.46 & 0.479 & 13.16\$ & 2.7h \\
        \hline
        $p_{as}$ & 146 & 123 & 166 & 143 & 0.54 & 0.543 & 0.505 & 0.523 & 5.11\$ & 0.7h \\
        \hline
        $p_{as-rci}$ & 131 & 59 & 230 & 158 & 0.625 & 0.689 & 0.453 & 0.547 & 18.48\$ & 2.4h \\
        \hline
        $p_{rf}$ & 161 & 168 & 121 & 128 & 0.488 & 0.489 & 0.557 & 0.521 & 11.86\$ & 6.2h \\
        \hline
        $p_{rf-rci}$ & 153 & 112 & 177 & 136 & 0.571 & 0.577 & 0.529 & 0.552 & 41.15\$ & 18h \\
        \hline
        $p_{fs20}$ & 147 & 150 & 139 & 142 & 0.495 & 0.495 & 0.509 & 0.502 & 17.39\$ & 0.4h \\
        \hline
        $p_{fs6}$ & 157 & 132 & 157 & 132 & 0.543 & 0.543 & 0.543 & 0.543 & 6.07\$ & 0.5h \\
        \hline
        $p_{fs6-rci}$ & 168 & 146 & 143 & 121 & 0.538 & 0.535 & 0.581 & 0.557 & 34.95\$ & 6.4h \\
        \hline
        $p_{dfa}$ & 150 & 102 & 187 & 139 & 0.583 & 0.595 & 0.519 & 0.555 & 9.33\$ & 4.8h \\
        \hline
        $p_{dfa-rci}$ & 171 & 57 & 232 & 118 & \textbf{0.697} & \textbf{0.75} & \textbf{0.592} & \textbf{0.662} & 34.58\$ & 20.5h \\
        \hline
        $p_{dfa-h}$ & 163 & 110 & 179 & 126 & 0.592 & 0.597 & 0.564 & 0.58 & 8.64\$ & 4.7h \\
        \hline
        $p_{dfa-h-rci}$ & 165 & 61 & 228 & 124 & 0.68 & 0.73 & 0.571 & 0.641 & 35.47\$ & 39.5h \\
        \hline
        $p_{cot-dfa}$ & 142 & 72 & 217 & 147 & 0.621 & 0.664 & 0.491 & 0.565 & 12.76\$ & 4.5h \\
        \hline
        $p_{cot-dfa-rci}$ & 146 & 65 & 224 & 143 & 0.64 & 0.692 & 0.505 & 0.584 & 43.61\$ & 11.2h \\
        \hline
        $p_{cot-8s}$ & 160 & 88 & 201 & 129 & 0.625 & 0.645 & 0.554 & 0.596 & 13.4\$ & 4.9h \\
        \hline
        $p_{cot-8s-rci}$ & 161 & 85 & 204 & 128 & 0.631 & 0.654 & 0.557 & 0.602 & 45.94\$ & 11.7h \\
        \hline
        $p_{cr}$ & 144 & 116 & 173 & 145 & 0.548 & 0.554 & 0.498 & 0.525 & 12.49\$ & 5.3h \\
        \hline
        $p_{cr-rci}$ & 142 & 110 & 179 & 147 & 0.555 & 0.563 & 0.491 & 0.525 & 43.83\$ & 12h \\
        \hline
    \end{tabular}
\end{table*}

Lastly, a prompting strategy called self-consistency is tested, which we denote as $p_{b-sc}$ \cite{self-consistency}. For this, the basic prompt is run three times, and only the files that are classified positive two or three times are counted as positive. This performs worse than the RCI, self-refinement and self-reflection strategies, most likely due to using a temperature value of zero. With higher temperature values, self-consistency could provide more benefits, as the higher temperature would cause the output to be more random, which the self-consistency strategy could help control.

\subsubsection{Comparing prompting approaches from prior studies}
While different prompting strategies have been proposed by prior studies, they have not been compared amongst each other on the same dataset. We try different approaches that have shown good results in previous studies and report the results to show which one performs the best on our dataset. 

Adding the API call sequence to the prompt along with the code has been shown to improve the LLM vulnerability detection capabilities \cite{vul-detect-prompt-enhanced}. To test that, API call sequence extraction capabilities are created, and the API sequence is provided to the LLM along with the code. This prompt is denoted as $p_{as}$ in Table \ref{tab:llm_results}, and it does outperform the basic prompting strategy. We also test the RCI approach on the results ($p_{as-rci}$), and while it does significantly lower the amount of false positive results, it also lowers the amount of true positive results. The RCI strategy does have an overall positive effect on the results, raising the accuracy, precision and F1 scores. Nonetheless, the basic prompting strategy benefits more from the RCI strategy and thus, the API sequence approach with the RCI strategy does not outperform the baseline basic prompt utilizing the RCI strategy.

It has been suggested that LLMs might be able to perform better if they are not only asked to find vulnerabilities but also to provide a fix for them \cite{noever2023-llm-fix}. To test that, we use a prompt that requires the LLM to detect vulnerabilities and to provide a fix for the found issue, denoted as $p_{rf}$ for required fix prompting in Table \ref{tab:llm_results}. The same prompt is tested utilising the RCI strategy as well, which we denote as $p_{rf-rci}$. Similarly to the API call sequence prompt, the requiring fix approach outperforms the basic prompt but sees very little improvement with the RCI strategy. Overall, this approach does not outperform the baseline basic prompt utilizing the RCI strategy. Additionally, the fix prompting strategy ends up being among the most costly strategies we test. This is because the output tokens are expensive, and the output is expected to contain the fixed code.

GPT-4 has shown good results in vulnerability detection tasks in few-shot settings. There are many potential ways to do few-shot prompting: using examples with different vulnerabilities, adjusting the number of examples, and modifying the balance between vulnerable and non-vulnerable samples. We follow the previous study, utilising examples provided by MITRE and including examples from the top 25 vulnerabilities \cite{llms-emerge-vuln-detect}. We have an equal number of vulnerable and non-vulnerable samples, and we try to use different sample sizes. First, we include all CWEs from the MITRE top 25, where they have a Java code example available. Overall, there are ten such CWEs, and as we also include a fixed version of all the samples, we end up with 20 examples overall. Thus we name this prompt $p_{fs20}$ in Table \ref{tab:llm_results}. Of the ten included CWEs, only two exactly match the CWEs in our dataset. To see how providing fewer examples affects the results, we also try a prompt containing six simple samples overall, three vulnerable and three non-vulnerable, which we call $p_{fs6}$. There we use one CWE, which is also present in our dataset and two that are not. This approach yields better results, most likely due to having less distracting code samples in the context. We can slightly improve the F1 score with the RCI strategy ($p_{fs6-rci}$); however, this raises not only the number of true positives but also the false positives. 

Asking LLMs to analyse the data flow has been shown to improve the vulnerability detection results \cite{understanding-llm-effectiveness}. This is similar to CoT prompting, as the LLM is asked to analyse the dataflows and sanitisers, but without explicitly stating to think step-by-step. To test that approach, we run a dataflow analysis prompt, denoted as $p_{dfa}$ in Table \ref{tab:llm_results}, similarly to \cite{understanding-llm-effectiveness}. While the given paper uses a self-refinement strategy to improve the results \cite{understanding-llm-effectiveness}, we use the RCI approach ($p_{dfa-rci}$), as we saw it perform better on the baseline results. The main difference between our approach and that of the original paper \cite{understanding-llm-effectiveness} is that they used LLMs for binary classification, whereas we use them for multi-class classification. The dataflow analysis prompt with RCI stands out as the best-performing approach. It outperforms the previous best approaches for the accuracy, precision, recall and F1 scores. What is more, this approach also outperforms the best CodeQL and SpotBugs results. While it is not the costliest strategy we test, the main downside of this strategy is that it is rather costly to run.

\subsubsection{Custom prompting strategies}
Testing the approaches proposed in previous studies allowed us to get good results. To see if these results could be further improved, multiple additional approaches are tested. To provide a better overview, the best results from the baseline testing and the previous studies approaches have been provided in Table \ref{tab:llm_results} among the new results.

As the dataflow analysis prompt with RCI ($p_{dfa-rci}$) shows the best results, modifications of this prompt are tested to improve the results. First, some more hints are added to the prompt, which we denote as $p_{dfa-h}$. This is a small change, where the name of the programming language (Java) is added, and the wording is changed to explicitly mention the file might not contain any vulnerabilities at all. This shows slight improvements over the original prompt ($p_{dfa}$). However, testing this approach with RCI strategy ($p_{dfa-h-rci}$) does not bring as big improvements as the original. While we manage to achieve an F1 score similar to the one of the original dataflow analysis with RCI, we are unable to improve the results. To see if asking the LLM to think step-by-step offers any improvements, the dataflow analysis prompt is modified into a CoT prompt, denoted as $p_{cot-dfa}$. Just as with adding the hints, we can get some improvements before the RCI strategy is applied. However, the RCI strategy ($p_{cot-dfa-rci}$) does not offer as many improvements as it did for the original dataflow analysis prompt.

A separate CoT strategy is developed to more closely mimic what software engineers would manually check during the code review process. This consists of eight steps, thus the prompt is named CoT eight-step, denoted as $p_{cot-8s}$. 
We ask the LLM to first identify all vulnerabilities that could be present in the given code. As the second step, we ask the LLM to review user input handling. The third step is to analyse the data flows. The fourth step checks for mitigations, and the fifth step evaluates conditional branching in the code. The sixth step is to assess error handling. The seventh step is to identify whether the code contains plaintext secrets. The last step is to provide a verdict. The exact wording of the dataflow and CoT 8-step prompts are given in ~\nameref{appendix:prompts}.
For the previous CoT approach we tested, we see very small improvements with the RCI strategy. In this case, the RCI once again has a very small impact on the F1 score. This is likely due to the CoT approach already enriching the context enough for the LLM, which means the RCI strategy does not add anything more meaningful. However, the CoT eight-step can produce fourth-best results overall with far smaller costs than those of the better-performing strategies.

Treating the exercise as code review seemed to perform rather well. Motivated by that observation, another prompt for a similar approach is developed. This time the LLM is given a checklist of questions and asked to treat this as a code review exercise denoted as $p_{cr}$ for the code review prompt. These questions contain quite generic questions that software engineers should think about when reviewing the code. The checklist was taken from the internet and slightly supplemented \cite{security-code-review-checklist}. This approach performs much poorer than the CoT 8-step approach in accuracy, precision, recall and F1 scores. Interestingly, this is the only approach which did not benefit from using the RCI strategy ($p_{cr-rci}$), showing the same F1 score both before and after using the RCI strategy.

\subsubsection{Strategies requiring higher temperature values}
Some prompting strategies benefit from using higher temperature values. We see that with the temperature value set to zero, the self-consistency approach does not provide many benefits. That makes sense, as the idea of self-consistency is to be able to get consistent responses with higher temperatures \cite{self-consistency}. Another strategy that should benefit from higher temperature values is the tree of thoughts (ToT) \cite{tot-prompt}. ToT strategy is somewhat similar to the CoT; however, for every step, multiple potential responses are generated. These responses are then evaluated, and the best one is chosen. As both the self-consistency and ToT have shown good results with temperature values of 0.7, that is the temperature we use \cite{tot-prompt, self-consistency}. 

To test these methods, we use the CoT 8-step prompt, as it provides the best results before applying any further steps (like the RCI strategy). As running these strategies is expensive, we start with testing a few select CWEs to better understand if they could outperform the previous attempts. Firstly, the CWE-23 is chosen because it has a very good detection rate and a plain CoT 8-step prompt. Secondly, CWE-129 is chosen, which has an average detection rate when using a plain CoT 8-step prompt. Thirdly, CWE-549 is chosen, which is not detected at all by the basic CoT 8-step strategy.

For the ToT strategy, we generate three potential candidates for each step and then have three different evaluators evaluate the responses for each step. The response chosen by the most evaluators is picked. This is similar to the approach the authors of the ToT paper utilised in the creative writing task \cite{tot-prompt}. For the testing, we forked the repository created by the authors of the ToT paper\footnote{\url{https://github.com/princeton-nlp/tree-of-thought-llm}} and added a task for code analysis\footnote{https://github.com/katamb/tree-of-thought-llm-ca}. For the strategy to work properly, we need to make slight modifications to the prompt and we need to have another prompt for the evaluation step. We call the main prompt $p_{tot-8s}$ and the evaluation prompt $p_{tot-8s-eval}$. In our case, as the strategy had eight steps, we generated three candidates for each step and had three evaluators. This means we made $8 \times (3+3) = 48$ calls to the LLM just for analysing a single file. As displayed in Table \ref{tab:3cwe_test_higher_temp}, for the evaluation of CWE-129, the strategy does provide slight improvements. However, for CWE-23, the results are significantly worse. Based on these results and the high costs, we do not dive deeper into testing the ToT strategy. It must be noted that this strategy has potential for future research, as there are many things to configure, like the evaluation strategy, the number of responses generated, etc.

\begin{table*}[htbp]
    \renewcommand{\arraystretch}{1.1}
    \caption{\it{Results from initial testing of higher-temperature strategies}}
    \label{tab:3cwe_test_higher_temp}
    \centering
    \begin{tabular}{|c||c||c|c|c|c||c|c|c|c||c|c|}
        \hline
        \textbf{Strategy} & \textbf{CWE} & \textbf{TP} & \textbf{FP} & \textbf{TN} & \textbf{FN} & \textbf{Accuracy} & \textbf{Precision} & \textbf{Recall} & \textbf{F1} & \textbf{Cost} & \textbf{Time} \\
        \hline
         & CWE-23 & 17 & 2 & 15 & 0 & 0.941 & 0.895 & 1 & 0.944 & & \\*
        \cline{2-10}
        $p_{cot-8s}$  & CWE-129 & 14 & 15 & 2 & 3 & 0.471 & 0.483 & 0.824 & 0.609 & 2.38\$ & 0.8h \\*
        \cline{2-10}
        $t=0$ & CWE-549 & 0 & 0 & 17 & 17 & 0.5 & 0 & 0 & 0 &  & \\
        \hline
         & CWE-23 & 17 & 1 & 16 & 0 & 0.971 & 0.944 & 1 & 0.971 & & \\*
        \cline{2-10}
        $p_{cot-8s-sc}$ & CWE-129 & 14 & 14 & 3 & 3 & 0.5 & 0.5 & 0.824 & 0.622 & 7.31\$ & 2.5h \\*
        \cline{2-10}
        $t=0.7$ & CWE-549 & 0 & 0 & 17 & 17 & 0.5 & 0 & 0 & 0 &  & \\
        \hline
         & CWE-23 & 17 & 11 & 6 & 0 & 0.676 & 0.607 & 1 & 0.756 & & \\*
        \cline{2-10}
        $p_{tot-8s}$ & CWE-129 & 11 & 5 & 12 & 6 & 0.676 & 0.688 & 0.647 & 0.667 & 49.08\$ & 5.9h \\*
        \cline{2-10}
         $t=0.7$ & CWE-549 & 0 & 0 & 17 & 17 & 0.5 & 0 & 0 & 0 &  & \\
        \hline
    \end{tabular}
\end{table*}

The self-consistency approach for the CoT 8-step prompt is denoted as $p_{cot-8s-sc}$. We run the same prompt three times and only count the classifications positive when the file is classified as positive two or three times. In Table \ref{tab:3cwe_test_higher_temp}, we can see the self-consistency approach with higher temperatures provides slight improvements for two of the three CWEs. As self-consistency shows improved results, we try the self-consistency strategy for the whole dataset. This resulted in a noticeable improvement in the results, especially regarding lowering the false positives. The results are displayed in Table \ref{tab:self_consistency_results}.

\begin{table*}[htbp]
    \renewcommand{\arraystretch}{1.1}
    \caption{\it{Self-consistency results}}
    \label{tab:self_consistency_results}
    \centering
    \begin{tabular}{|c||c|c|c|c||c|c|c|c||c|c|}
        \hline
        \textbf{Strategy} & \textbf{TP} & \textbf{FP} & \textbf{TN} & \textbf{FN} & \textbf{Accuracy} & \textbf{Precision} & \textbf{Recall} & \textbf{F1} & \textbf{Cost} & \textbf{Time} \\
        \hline
        $p_{cot-8s}$, $t=0$ & 160 & 88 & 201 & 129 & 0.625 & 0.645 & 0.554 & 0.596 & 13.4\$ & 4.9h \\
        \hline
        $p_{cot-8s-rci}$, $t=0$ & 161 & 85 & 204 & 128 & 0.631 & 0.654 & 0.557 & 0.602 & 45.94\$ & 11.7h \\
        \hline
        $p_{cot-8s-sc}$, $t=0.7$ & 164 & 60 & 229 & 125 & 0.68 & 0.732 & 0.567 & 0.639 & 40.81\$ & 15.4h \\
        \hline
    \end{tabular}
\end{table*}

\subsubsection{Different models}
As all the testing so far has been conducted on the GPT-4 turbo model, we also wanted to see how it compares to other commercial LLMs that are beloved by the users \cite{chatbot-leaderboard}. At the time of writing, the GPT-4 non-turbo model and Claude 3 Opus models are among the highest-ranking ones, so we also test them. The Google Gemini model is also considered. However, as it is not officially available in Europe\footnote{\url{https://ai.google.dev/available\_regions}} at the time of running the experiments, it is not included. For testing with other models, the temperature value of 0 is used.

Both these models are more expensive to run, so we only run a few strategies that show good results with the GPT-4 turbo model. We choose the dataflow analysis prompt, the dataflow analysis prompt with RCI and the CoT 8-step prompt. The dataflow prompt with the RCI re-evaluation was the best-performing one and the CoT 8-step prompt was the best-performing strategy without any re-evaluation steps. 

The results are displayed in Table \ref{tab:other_models_results}. Interestingly, both of the other models show better results with the CoT 8-step prompt. The GPT-4 model with the CoT 8-step prompt outperforms all the other approaches, showing F1 scores of 0.672. The results of the Claude 3 Opus model are also intriguing. The CoT 8-step strategy shows the lowest false positive rate we have seen among the LLMs whilst maintaining a respectable true positive rate. What is more, the RCI strategy improves the scores in all tests conducted on the GPT-4 turbo model. On the Claude 3 Opus model, the RCI strategy seems to reduce the performance.

\begin{table*}[htbp]
    \renewcommand{\arraystretch}{1.4}
    \caption{\it{Other models results}}
    \label{tab:other_models_results}
    \centering
    \begin{tabular}{|c||c||c|c|c|c||c|c|c|c||c|c|}
        \hline
        & \textbf{Strategy} & \textbf{TP} & \textbf{FP} & \textbf{TN} & \textbf{FN} & \textbf{Accuracy} & \textbf{Precision} & \textbf{Recall} & \textbf{F1} & \textbf{Cost} & \textbf{Time} \\
        \hline
        \parbox[t]{2mm}{\multirow{3}{*}{\rotatebox[origin=c]{90}{\scalebox{0.75}{GPT-4 turbo}}}} & $p_{dfa}$ & 150 & 102 & 187 & 139 & 0.583 & 0.595 & 0.519 & 0.555 & 9.33\$ & 4.8h \\
        \cline{2-12}
        & $p_{dfa-rci}$ & 171 & 57 & 232 & 118 & 0.697 & 0.75 & 0.592 & 0.662 & 34.58\$ & 20.5h \\
        \cline{2-12}
        & $p_{cot-8s}$ & 160 & 88 & 201 & 129 & 0.625 & 0.645 & 0.554 & 0.596 & 13.4\$ & 4.9h \\
        \hline
        \parbox[t]{2mm}{\multirow{3}{*}{\rotatebox[origin=c]{90}{\scalebox{0.75}{GPT-4}}}} 
        & $p_{dfa}$ & 154 & 98 & 191 & 135 & 0.597 & 0.611 & 0.533 & 0.569 & 19.35\$ & 1.5h \\
        \cline{2-12}
        & $p_{dfa-rci}$ & 148 & 37 & 252 & 141 & 0.692 & 0.8 & 0.512 & 0.624 & 66.79\$ & 3.6h \\
        \cline{2-12}
        & $p_{cot-8s}$ & 174 & 55 & 234 & 115 & \textbf{0.706} & 0.76 & \textbf{0.602} & \textbf{0.672} & 23.32\$ & 2h \\
        \hline
        \parbox[t]{2mm}{\multirow{3}{*}{\rotatebox[origin=c]{90}{\scalebox{0.75}{Claude 3 Opus}}}} 
        & $p_{dfa}$ & 141 & 71 & 218 & 148 & 0.621 & 0.665 & 0.488 & 0.563 & 19.32\$ & 3.1h \\
        \cline{2-12}
        & $p_{dfa-rci}$ & 112 & 50 & 239 & 177 & 0.607 & 0.691 & 0.388 & 0.497 & 65.29\$ & 10.8h \\
        \cline{2-12}
        & $p_{cot-8s}$ & 137 & 18 & 271 & 152 & \textbf{0.706} & \textbf{0.884} & 0.474 & 0.617 & 26.87\$ & 3.8h \\
        \hline
    \end{tabular}
\end{table*}

\subsection{Overall Comparative Analysis}
We demonstrate the best-performing results in all models with baseline static tool analysis results in Table \ref{tab:results_overview}. The best overall precision is achieved by CodeQL with its default configuration, achieving a precision score of 0.938. The best overall results from traditional static analysis tools are shown by SpotBugs with the Find Security Bugs plugin. This shows the best overall accuracy score of 0.723 while maintaining respectable precision, recall and F1 scores. We provide the best prompting strategy for each of the tested LLMs. The GPT-4 turbo performs best with the $p_{dfa-rci}$ prompt, outperforming the static analysis tools in terms of F1 score. The GPT-4 performs best with the $p_{cot-8s}$ prompt, showing the best recall and F1 scores overall. The Claude 3 Opus model performs best with the $p_{cot-8s}$ prompt, showing the lowest false positive rate out of the tested LLM approaches. In terms of price and time, the traditional tools provide a much better value. 

\begin{table*}[htbp]
    \renewcommand{\arraystretch}{1.2}
    \caption{\it{Results overview}}
    \label{tab:results_overview}
    \centering
    \begin{tabular}{|c||c|c|c|c||c|c|c|c||c|c|}
        \hline
        & \textbf{TP} & \textbf{FP} & \textbf{TN} & \textbf{FN} & \textbf{Accuracy} & \textbf{Precision} & \textbf{Recall} & \textbf{F1} & \textbf{Cost} & \textbf{Time} \\
        \hline
        CodeQL-d & 76 & 5 & 284 & 213 & 0.623 & \textbf{0.938} & 0.263 & 0.411 & 0\$ & \textless{1m} \\
        \hline
        CodeQL-esq & 137 & 23 & 266 & 152 & 0.697 & 0.856 & 0.474 & 0.61 & 0\$ & \textless{1m} \\
        \hline
        SpotBugs-fsb & 152 & 23 & 266 & 137 & \textbf{0.723} & 0.869 & 0.526 & 0.655 & 0\$ & \textless{1m} \\
        \hline
        GPT-4 turbo & \multirow{2}{*}{171} & \multirow{2}{*}{57} & \multirow{2}{*}{232} & \multirow{2}{*}{118} & \multirow{2}{*}{0.697} & \multirow{2}{*}{0.75} & \multirow{2}{*}{0.592} & \multirow{2}{*}{0.662} & \multirow{2}{*}{34.58\$} & \multirow{2}{*}{20.5h} \\*
        $p_{dfa-rci}$ &  &  &  &  &  &  &  &  &  & \\
        \hline
        GPT-4 & \multirow{2}{*}{174} & \multirow{2}{*}{55} & \multirow{2}{*}{234} & \multirow{2}{*}{115} & \multirow{2}{*}{0.706} & \multirow{2}{*}{0.76} & \multirow{2}{*}{\textbf{0.602}} & \multirow{2}{*}{\textbf{0.672}} & \multirow{2}{*}{23.32\$} & \multirow{2}{*}{2h} \\
        $p_{cot-8s}$ &  &  &  &  &  &  &  &  &  &  \\
        \hline
        Claude 3 Opus & \multirow{2}{*}{137} & \multirow{2}{*}{18} & \multirow{2}{*}{271} & \multirow{2}{*}{152} & \multirow{2}{*}{0.706} & \multirow{2}{*}{0.884} & \multirow{2}{*}{0.474} & \multirow{2}{*}{0.617} & \multirow{2}{*}{26.87\$} & \multirow{2}{*}{3.8h} \\*
        $p_{cot-8s}$ &  &  &  &  &  &  &  &  & & \\
        \hline
    \end{tabular}
\end{table*}

From the 17 CWEs in our dataset, CodeQL is unable to detect any vulnerabilities for six CWEs, namely CWE-81, CWE-256, CWE-523, CWE-549, CWE-566 and CWE-606. From these six, two CWEs, namely CWE-566 and CWE-606, are not supported by the CodeQL mapping strategy \cite{github-codeql-cwe-mapping}. This means CodeQL has no mapping strategy to detect the CWE itself nor any of the parent or child CWEs. From the 17 CWEs in our dataset, SpotBugs is unable to detect any vulnerabilities for six CWEs, namely CWE-190, CWE-256, CWE-523, CWE-549, CWE-566 and CWE-606. Five of these are the same CWE categories that CodeQL is unable to detect. Interestingly, CodeQL is unable to detect any vulnerabilities for the CWE-81, whereas SpotBugs achieves perfect results with that CWE category. However, CodeQL can correctly identify seven out of the 17 CWE-190 vulnerabilities, whereas SpotBugs finds none. The GPT-4 model using the CoT 8-step prompt and the GPT-4 turbo model using the dataflow prompt with RCI strategy are both unable to correctly identify any true positives for four out of the 17 CWEs: CWE-523, CWE-549, CWE-566 and CWE-606. The Claude 3 Opus model performs the best by only failing to identify any true positives for three CWE categories: CWE-523, CWE-566 and CWE-606.

\subsection{Qualitative analysis}
For the vulnerability detection tools to provide value, they must not only point at a vulnerability but also explain their findings. We compare the outputs from CodeQL, SpotBugs and two LLM model outputs. The best-performing approach for each tool is used for the comparison. For CodeQL, the output of CodeQL-esq configuration is used, for SpotBugs, the output of SpotBugs-fsb configuration is used. For LLM, the GPT-4 and Claude 3 Opus models with CoT 8-step prompting results are used. The choices of LLMs were based on the best overall recall, precision and F1 scores. The exact outputs of different tools are given in ~\nameref{appendix:qualitative-analysis}.

The file named "J20736" in the dataset is vulnerable to CWE-78: OS command injection. All approaches we cover correctly identify the file as vulnerable. The vulnerable function assigns the variable "data" a user-provided value, which is then used to execute the system command. The CodeQL-esq scan reports the issue correctly. The problem is reported as an "Uncontrolled command line", with an easily understandable description and line numbers. The result contains the name, description, severity, message, path, start line, start column, end line and end column of the problem \cite{codeql-csv-explanations}. The SpotBugs-fsb scan correctly identifies the issue and the line number. The problem is classified as a high-severity security issue. The description of the problem, the file name and the line number are provided. The GPT-4 LLM (using $p_{cot-8s}$) correctly identifies the relevant issue. The verdict contains the correct CWE identifier and the correct description. The description is concise and easy to follow in a nice human-readable format. The Claude 3 Opus LLM (using $p_{cot-8s}$) also finds the issue and provides just as nice a description as the GPT-4 model. Overall, the file is correctly identified as vulnerable to OS command injection by all four approaches. All four outputs are clear on the vulnerability type, with the LLM response being the most verbose.

The file named "J23877" in the dataset is not vulnerable to CWE-89: SQL injection. However, all the approaches incorrectly identify the file to be vulnerable. The file contains two non-vulnerable functions, one with a good source and a bad sink, and the other with a bad source and a good sink. While these functions do not follow the best practices, neither function can be exploited. The CodeQL-esq scan reports two issues related to SQL injection. The problem we are interested in is reported as a "Query built by concatenation with a possibly untrusted string". The description states that the value used in the query "may be untrusted". While it is indeed correct that using string concatenation is not the best practice for SQL queries, in this case, the code is not vulnerable, as the variable can not be set by the user. The result is still counted as a false positive, as the problem is reported as SQL injection instead of not following the best practices. The SpotBugs-fsb reports one SQL injection-related issue. The issue is reported as a medium-level security issue. The wording hints at a possible SQL injection, which in the given case can not happen and thus is counted as a false positive. The GPT-4 LLM (using $p_{cot-8s}$) marks the function with the bad source to be vulnerable, as the code does not use a prepared statement for the SQL query. The analysis fails to notice the value concatenated to the SQL statement is hard-coded and thus is not considered a vulnerability. The Claude 3 Opus LLM (using $p_{cot-8s}$) also marks the function with the bad source to be vulnerable, as the code does not use a prepared statement for the SQL query. While the output correctly points out the code is currently not exploitable, it still reports "vulnerability: YES $\mid$ vulnerability type: CWE-89", which we consider a positive classification. Overall, all approaches incorrectly identify an SQL injection vulnerability in the code. The wording of the problem in the case of CodeQL and SpotBugs hints that string concatenation should not be used in SQL statements. Similarly, Claude 3 Opus model correctly mentions that string concatenation should not be used. GPT-4 incorrectly states the function to be vulnerable to SQL injection. We classify all results as false positives, as the code is not exploitable, but all approaches report SQL injection vulnerabilities. The Claude 3 Opus model provides the best verdict, correctly noticing that the code is currently not exploitable.

\section{Discussion}\label{_Discussion_}
\subsection{What prompting approach is most successful with the LLMs to detect vulnerabilities?}
Many different prompting strategies have been proposed by previous studies for vulnerability detection with LLMs \cite{understanding-llm-effectiveness, vul-detect-prompt-enhanced, noever2023-llm-fix, llms-emerge-vuln-detect}. The studies use different datasets, they try to detect different CWEs, and some frame the problem as binary classification, others as a multi-class classification task. Compiling a dataset of 578 Java files covering 17 different CWEs allows us to compare different prompting strategies and models to see which performs the best for multi-class classification tasks. The reason for testing the multi-class classification capabilities is that this provides more value in the real-world setting. The traditional static analysis tools are even more user-friendly, providing the exact line numbers for the vulnerability, which is called fine-grained classification. We believe LLMs come close to that ability, as in addition to detecting the problems, they also explain them. However, we do not ask for exact line numbers to simplify the evaluation of the correctness of the responses.

Most prior studies utilise GPT-3.5 and GPT-4 models, with the latter always outperforming the former in vulnerability detection tasks \cite{chatgpt-devsecops, gpt-vul-detect-cass-rep, gpt-source-code-inspection}. Unfortunately, in most cases, the exact versions of the LLMs are not provided, which makes it difficult to compare the results. After establishing the best-performing prompt from previous studies, we suggest other prompting strategies based on the research in the prompt engineering field. We try prompting approaches that have shown promise in other domains and adapt them for vulnerability detection tasks. We test the CoT \cite{chain-of-thought} and different self-refinement strategies \cite{self-refinement-prompting}. What is more, we do some testing with prompting approaches requiring higher temperature values like self-consistency \cite{self-consistency}, and ToT \cite{tot-prompt}. Finally, we test the most promising approaches on more expensive models, where the best overall results are shown with a CoT 8-step prompt devised by us. To the best of our knowledge, ToT and self-consistency have not been tested for vulnerability detection tasks before. Some self-refinement strategies have been tested before \cite{understanding-llm-effectiveness}, but we try different approaches for self-refinement \cite{recursive-criticism-and-improvement, understanding-llm-effectiveness, self-refinement-prompting} and find RCI strategy works best for our use case. While CoT prompting has been suggested before for vulnerability detection tasks \cite{vul-detect-prompt-enhanced}, these efforts are very different from ours. The CoT prompt proposed previously considers a two-step approach: the first step is to explain the code, and the second is to find issues with the code \cite{vul-detect-prompt-enhanced}. The CoT 8-step prompt we propose lists eight steps that should be taken to find potential issues in source code. While this approach does not outperform the dataflow analysis prompt on the GPT-4 turbo model, it shows the best results on other models we test. This highlights that different prompting techniques can have advantages on different models and that the prompt should be adapted to the LLM used.

The best results achieved with LLMs slightly outperform CodeQL and SpotBugs analysis results. Depending on the LLM used, different prompting approaches can be more successful. We suggest using the CoT 8-step prompt proposed by us with either GPT-4 or Claude 3 Opus model. In our case, half the files in the dataset are vulnerable, whereas in the real world, there will likely be multiple non-vulnerable files for every vulnerable one. This is also likely one of the reasons why, for example, the CodeQL default configuration had a low recall but high precision value. High precision means fewer false alarms and better user experience. Industry experts seem to place a high value on having as few false positives as possible, with Google noting false positives should make up less than 10\% of all reported issues \cite{google-sat-lessons}.

If the expectation is to find as many vulnerabilities as possible and the amount of false positive results is not as important, then we suggest using the GPT-4 model. This approach has the best recall and F1 scores, which means on paper, this is the best approach. If the expectation is to get as few false positive results as possible, we suggest using the Claude 3 Opus model. Claude 3 Opus model produces impressively few false positive results and could thus be the most pleasant to use in real-world scenarios. This is also the only approach that gets close to the 10\% figure ($\frac{18 \times 100}{155} = 11.6\%$) of false positives that Google expects \cite{google-sat-lessons}. The GPT-4 turbo model using the dataflow analysis prompt with the RCI strategy also performs well. Although the cost per token for the GPT-4 turbo model is significantly cheaper, the RCI strategy adds extra complexity, requiring more tokens. This makes the best-performing strategy on the GPT-4 turbo model more expensive than the better-performing GPT-4 or Claude 3 Opus models. Thus we recommend using the CoT eight-step prompt in conjunction with GPT-4 and Claude 3 Opus models for vulnerability detection tasks.

\subsection{What advantages and disadvantages do LLMs have over existing static analysis tools?}
The most important disadvantage of LLMs over existing static analysis tools is the cost, both monetary and time-wise. While the high cost of LLMs is mentioned often when discussing LLMs for vulnerability detection, to the best of our knowledge, the cost of different prompting approaches has not been discussed in detail. We provide details on the costs for every prompting strategy, allowing for comparison by not only the performance but also cost. The cost factor is considered when making recommendations and should be considered when opting to use commercial LLMs in vulnerability detection tasks. Depending on the dataset, different approaches could make sense. If the dataset is large, some prompting strategies, like ToT, become impractical due to their high costs. We also find that even though the RCI strategy usually helps to improve the results of OpenAI models, it significantly increases the costs, usually by a factor of three or more.

The CodeQL and SpotBugs are both easy to set up and use. Both tools are free to use, and the analysis of our dataset can be completed in under a minute. This means they are orders of magnitude faster than the fastest LLM-based approaches. Even though the LLM-based approach slightly outperforms the traditional tools based on the F1 score, the difference is rather small. The main problem with the LLM-based approach is the time it takes to run and the monetary cost associated. The biggest strength of the LLM-based analysis is the ability to detect most vulnerabilities. 

Both the static analysis tools completely miss six vulnerability classes from the 17 in our dataset. For LLMs, most analyses missed four CWE classes, with the Claude 3 Opus not having any correct positive classifications on only three of the CWE classes. Analysing the LLM responses manually, we see that the descriptions often point out the correct issue in the code but fail to associate the correct CWE to it. The first CWE missed by all LLM analyses is the CWE-523, which, in the case of our dataset, means that HTTP protocol is used where HTTPS protocol should be used. Both suggested approaches describe the issue correctly on multiple occasions but provide a CWE identifier, which we consider incorrect. For example, the LLMs often classify the problem as CWE-319, which points to cleartext transmission of sensitive data. While the issues are similar, the matching strategy we use allows us to count only the parents and children of the expected CWE as correct. Based on the CWE research view (CWE-1000), CWE-319 is neither a child nor a parent of CWE-523, meaning we do not count it as a positive classification. The second CWE missed by most LLMs is the CWE-549. In our dataset, this problem exists due to Java code generating an HTML form with a password field, where the field is marked as “text” instead of the expected “password” type. While this problem is correctly classified once, in most cases, the description matches the problem, but the CWE identifier provided by LLM is not considered correct. Similarly to CWE-523, the CWE-319 is often offered as a CWE identifier when we expect CWE-522 or CWE-523. The third CWE missed by all LLM analyses is the CWE-566. The CWE-566 means the code uses the user-provided value as the primary key in querying the database without checking the user access rights. While this is a valid vulnerability, correctly identifying the issue often requires knowledge of the business context. The LLMs, at times, do mention the user-provided value is used in the SQL query, but the issue is classified as CWE-89: SQL injection. Classifying the issue as SQL injection can be confusing, as prepared statements are used and whether the code is vulnerable depends on the context. The fourth CWE missed by all LLM analyses is the CWE-606, which is caused by using user inputs in loop conditions without proper validation. The LLMs are also able to point at the problem correctly in many instances but identify the issue as CWE-400, which points to uncontrolled resource consumption. Overall, we find the LLMs can correctly describe the problems for 16 CWE categories out of the 17 in our dataset. For comparison, CodeQL and SpotBugs can detect 11 CWE categories out of the 17. This shows the LLMs have an advantage in finding vulnerabilities not supported by traditional tools.

\subsection{How would the use of off-the-shelf LLMs be able to contribute to vulnerability detection in source code?}
While the LLMs show good ability in detecting many different types of vulnerabilities, they are significantly more expensive and time-consuming to run than traditional static analysis tools. Thus we believe that static code analysers like CodeQL and SpotBugs are currently still better for everyday use for developers. They run quickly, are easy to set up and are free to use. They are configured to produce rather few false positives and provide quite good descriptions of problems together with line numbers. However, using LLMs for security analysis could be justified in some cases, like during security audits. The LLMs show the ability to find a larger variety of issues and explain them rather well. Another advantage of using an LLM-based approach includes not needing the code to be compileable. This allows for an easy analysis of some parts of the code without providing access to the full codebase.

\subsection{Other lessons learned}
Previous studies mostly utilise different OpenAI models, like GPT-3.5 and GPT-4 \cite{understanding-llm-effectiveness, vul-detect-prompt-enhanced, cot-vsp-llms, chatgpt-devsecops, java-gpt-eval, gpt-vul-detect-cass-rep, gpt-source-code-inspection}. Some other studies also incorporate open-source models from the Llama \cite{understanding-llm-effectiveness, cot-vsp-llms}, BERT \cite{llms-emerge-vuln-detect, gpt-effectiveness} and Falcon series \cite{cot-vsp-llms}, with one study also including Google's Gemini model \cite{sec-code-review-llm-responses}. To the best of our knowledge, we are the first to include the Claude 3 Opus model from Anthropic in the comparisons.

The cost structure of commercial LLM API usage is similar between models, requiring users to pay for input and output tokens. Output tokens are significantly more expensive. From the cost perspective, it is reasonable to add as much context into the input and ask the LLM to provide short answers. To save costs and to be able to evaluate the analysis results automatically, we specify the format in which to provide the final verdict. The format we use is: "vulnerability: \textless{YES or NO}\textgreater $\mid$ vulnerability type: \textless{CWE\_ID}\textgreater $\mid$". Usually, the LLMs follow the provided format, but not always. Interestingly, different models have different deviations from the provided format.
The GPT-4 turbo model often adds decorators like: "vulnerability: **YES** $\mid$ vulnerability type: **CWE-89**" or provides a vulnerability description instead of the keyword "vulnerability". The GPT-4 model sometimes provides the CWE identifier with an underscore: "CWE\_89". The Claude 3 Opus model often uses line changes instead of the "$\mid$" sign. Overall, for better-performing strategies, we manually inspect the responses that contain the word "YES" and make sure the format of the response does not affect the results. There are some rare cases where the model provided "vulnerability: MAYBE" or "vulnerability: POSSIBLE" instead of "YES" or "NO". As we know that the context needed for discovering the vulnerability should be given in the file, we count these as "NO". There are very few such classifications, and in most cases, they refer to irrelevant CWEs. The testing we conduct with higher temperature values utilising self-consistency and ToT strategies has a lot more responses that do not adhere to the format. This is likely related to the temperature increase, which makes the responses more random and affects the output format. This means the testing with higher temperature values would require either a better automatic mapping strategy or more manual effort to benchmark.

\subsection{Limitations and threats to validity}
The dataset is synthetic and the multi-class vulnerability detection is tested on file level, not line or function level (not fine-grained classification). The performance on real-world datasets or for fine-grained classification may vary. We use a dataset where the vulnerabilities can be detected based on a single file. The detection capabilities of any of the tested tools might not directly translate to more complex real-world projects. In our case, half the files are vulnerable, and the other half are not. In real-world projects, most files do not contain vulnerabilities. Even the well-established static analysis tools, which scan the whole codebase at once, have been shown to perform significantly worse on real-world datasets when compared to synthetic ones \cite{sast-comparison}. However, there have been studies using LLMs for vulnerability detection using real-world datasets, which have shown good results \cite{llms-emerge-vuln-detect, llm4vuln}. What is more, it is likely the capabilities of the LLMs will improve for large, more complex codebases, where multiple files need to be analysed at once. The token limits have been increasing rapidly in the past years, with Google announcing having successfully tested context windows of up to ten million tokens \cite{google-gemini-15}. The larger context window can help translate the results seen on synthetic datasets to real-world projects.

17 unique CWE categories are covered, which map to 11 CWE categories from MITRE's top 25 list. While this might not cover all important vulnerability categories, it does cover a large portion of what has been classified as the most dangerous by MITRE \cite{mitre-cwe-top-25-methodology}. We only focus on Java programming language; however, based on prior research, the approaches likely translate reasonably well to other popular programming languages \cite{understanding-llm-effectiveness}. Our contribution includes benchmarking on a unique dataset containing more CWEs than most previous studies. Most prior studies have focused on five or fewer CWEs to evaluate the results \cite{understanding-llm-effectiveness, cot-vsp-llms, java-gpt-eval, gpt-source-code-inspection}.

The Juliet dataset might be present in the training data of the LLM, which could affect the results. To mitigate that issue as much as possible, we conduct extensive pre-processing of the dataset. During pre-processing, file structure, file naming, and the names of variables and functions are changed. Furthermore, the package names are also hidden from the LLM and the indentation and format of the files are altered. These modifications should make it more difficult for the model to base the predictions on what has been present in the training data, as the files differ noticeably.

We employ a strategy for mapping CWEs which does not expect the CWE identifier to be identified exactly, similar to studies of static code analysers \cite{static-anal-cwe-grouping, static-anal-cwe-grouping-2}. By using the CWE-1000 research view, we can, in most cases, match the provided CWE to the expected one if the issue is identified correctly. However, we notice that this strategy does not work well for four of the CWEs in our dataset. Manually reviewing the LLM responses, we see correctly identified issues, which we can not automatically consider correct. 

Due to their high costs, we do not further evaluate the capabilities of GPT-4 and Claude 3 Opus models. The performance of these models could likely be further improved with higher temperatures utilising more elaborate prompting strategies, like self-consistency strategy, just like we saw improvements for the GPT-4 turbo model. To limit the scope, we focus on static code analysis. Incorporating other methods like dynamic analysis or other deep learning methods is out of the scope. Fine-tuning LLMs is out of scope, as the focus is on the prompting techniques. We recognise that the LLMs and the rulesets of the static code analysis tools can change over time. Thus the exact versions we used for experimentation are given in ~\nameref{appendix:versions}.

Data leakage to the LLM owners or hosting services should be considered before using commercial LLM APIs. Before sending any confidential data to any of the LLM providers, policies should be in place to ensure the safety of the confidential data. The OpenAI privacy policy states that by default, the data provided by the users of ChatGPT can be used to train the models, however, there is an opt-out option \cite{gpt-4-privacy-policy}.

\section{Conclusion}\label{_Conclusion_}
We consider the use of state-of-the-art LLMs for vulnerability detection tasks and compare the results with two traditional static code analysers. The purpose is to discover if LLMs could help in detecting vulnerabilities in source code. We are interested in whether LLMs have advantages or disadvantages over existing static code analysis tools. We run experiments and use comparative analysis techniques to evaluate the performance of different approaches. We consider different prompt engineering techniques previously not tested for vulnerability detection tasks. We find off-the-shelf LLMs show remarkable abilities in file-level vulnerability detection tasks. The success of a particular prompting strategy is dependent on the underlying LLM. The GPT-4 turbo model shows the best performance with dataflow analysis prompt utilising the RCI strategy. Meanwhile, GPT-4 and Claude 3 Opus models show better performance with a CoT 8-step prompt. The best prompting approaches outperform the static code analysis tools based on recall and F1 scores. The advantages of LLMs over static analysis tools include the ability to detect a larger variety of different vulnerabilities and a higher amount of true positive classifications. The disadvantages of LLMs include slower running time, higher costs, non-deterministic results and a higher amount of false positives. Thus we can show LLMs show a remarkable ability in vulnerability detection and multi-class classification tasks when all required information is provided in the context.

\section{Future Work}\label{_Future_Work_}
As long as the LLMs continue to improve and more capable models are released, the capabilities of these new models should be tested. A comparison of the performance and cost between commercial models and open-source models would be an interesting area to explore. Overall, more combinations of prompting strategies and LLMs should be tested to discover the best approaches. The ToT prompting strategy allows for many variations of different prompts and parameters, remaining a compelling area for future research. Fine-tuning the LLMs has been tried on some smaller models \cite{understanding-llm-effectiveness} and remains an interesting area for further exploration. 

Using other datasets for testing the proposed prompting approaches could provide valuable information on the performance of the prompts in different settings. Including more programming languages and testing the LLM capabilities for fine-grained vulnerability detection could be another vertical to explore. Future studies could further improve upon the CWE matching strategy to make sure the correctly described problems are counted as positive classifications without relying solely on the LLM to provide an acceptable CWE identifier.
The potential synergies between traditional tools and LLMs could be further researched. The LLMs have shown an ability to generate fixes for code \cite{noever2023-llm-fix} but the quality of these fixes has not been evaluated in detail. What is more, it would be intriguing to test the capabilities of LLMs for generating tests to prove the vulnerability is present. This could be done to reduce the amount of false positive classifications.

\addcontentsline{toc}{chapter}{References}
\printbibliography[title=References]

\clearpage

\section*{Appendix -- Prompts}\label{appendix:prompts}
Dataflow analysis prompt ($p_{dfa}$):
\begin{lstlisting}[ basicstyle=\scriptsize\ttfamily,
                    columns=fullflexible,
                    keepspaces=true,
                    breaklines=true,
                    frame=single,
                    framerule=0pt,
                    backgroundcolor=\color{gray!10}]
Human: You are a security researcher, expert in detecting security vulnerabilities. Carefully analyze the given code snippet and track the data flows from various sources to sinks. Assume that any call to an unknown external API is unsanitized.
    Please provide a response only in the following format:
    Here is a data flow analysis of the given code snippet:
    A. Sources: <numbered list of input sources>
    B. Sinks: <numbered list of output sinks>
    C. Sanitizers: <numbered list of sanitizers, if any>
    D. Unsanitized Data Flows: <numbered list of data flows that are not sanitized in the format (source, sink, why this flow could be vulnerable)>
    E. Vulnerability analysis verdict: vulnerability: <YES or NO> | vulnerability type: <CWE_ID> | vulnerability name: <NAME_OF_CWE> | explanation: <explanation for prediction>
    Is the following code snippet prone to any security vulnerability?
    ```{code}```
AI: <response>
\end{lstlisting}

Dataflow analysis prompt with RCI ($p_{dfa-rci}$):
\begin{lstlisting}[ basicstyle=\scriptsize\ttfamily,
                    columns=fullflexible,
                    keepspaces=true,
                    breaklines=true,
                    frame=single,
                    framerule=0pt,
                    backgroundcolor=\color{gray!10}]
Human: You are a security researcher, expert in detecting security vulnerabilities. Carefully analyze the given code snippet and track the data flows from various sources to sinks. Assume that any call to an unknown external API is unsanitized.
    Please provide a response only in the following format:
    Here is a data flow analysis of the given code snippet:
    A. Sources: <numbered list of input sources>
    B. Sinks: <numbered list of output sinks>
    C. Sanitizers: <numbered list of sanitizers, if any>
    D. Unsanitized Data Flows: <numbered list of data flows that are not sanitized in the format (source, sink, why this flow could be vulnerable)>
    E. Vulnerability analysis verdict: vulnerability: <YES or NO> | vulnerability type: <CWE_ID> | vulnerability name: <NAME_OF_CWE> | explanation: <explanation for prediction>
    Is the following code snippet prone to any security vulnerability?
    ```{code}```
AI: <response>
Human: Review your previous answer and find problems with that answer. Feel free to answer in any format you prefer.
AI: <response>
Human: Based on the problems found, improve your initial answer:
    Provide response only in following format:
    vulnerability: <YES or NO> | vulnerability type: <CWE ID> | vulnerability name: <CWE NAME> | explanation: <reason for the initial analysis being correct or wrong>.
    Use N/A in other fields if there are no vulnerabilities. Do not include anything else in response.
AI: <response>
\end{lstlisting}

CoT approach evaluating potential problems in eight steps ($p_{cot-8s}$):
\begin{lstlisting}[ basicstyle=\scriptsize\ttfamily,
                    columns=fullflexible,
                    keepspaces=true,
                    breaklines=true,
                    frame=single,
                    framerule=0pt,
                    backgroundcolor=\color{gray!10}]
Human: As a security researcher, you are tasked with reporting security vulnerabilities. You are currently looking at the following code:
    ```java
    {code}
    ```
    You are expected to work step-by-step:
    1. Identify Weaknesses: Identify which vulnerabilities could be present in the following code.
    2. Review User Input Handling: Look for any input sources that are not properly validated or sanitized before being used in unsafe manner. If variable that is passed into unsafe function is not directly influenced by external user input, the vulnerability is not currently present and must not be reported.
    3. Analyze Data Flow: Trace the flow of untrusted data to the system command. Ensure that there are no points where user-controlled input can directly influence the command execution.
    4. Check for Mitigations: Examine if there are any mitigations in place to prevent command injection, such as input validation, sanitization, or using safer alternatives to executing system commands.
    5. Evaluate Conditional Branching: If there's dead code (that can never be reached), then this part must not be evaluated.
    6. Assess Error Handling: Evaluate how errors, if any, are handled.
    7. Identify Code Leaking Secrets: Check whether the code contains secrets that should not be public knowledge.
    8. Provide verdict (one line for every potential discovered weakness). Keep in mind you must not report vulnerabilities that cannot be currently abused by malicious actors. False positive results must be kept to minimum. The verdict must be in the format:
    vulnerability: <YES or NO> | vulnerability type: <CWE_ID> |
    ...
AI: <response>
\end{lstlisting}

\section*{Appendix -- Versions}\label{appendix:versions}
\textbf{The following versions of different software are used in the study:}
\begin{enumerate}
    \item CodeQL: The CLI toolchain release 2.16.0 is used.
    \item SpotBugs: The Gradle plugin version 6.0.7 is used.
    \item FindSecurityBugs: The Find Sec Bugs plugin version 1.12.0 is used.
    \item GPT-4 turbo: The model gpt-4-0125-preview is used.
    \item GPT-4: The model gpt-4-0613 is used.
    \item Claude 3 Opus: The model claude-3-opus-20240229 is used.
    \item Juliet dataset: The Java Juliet 1.3 dataset is used.
\end{enumerate}

\section*{Appendix -- Qualitative Analysis}\label{appendix:qualitative-analysis}
The CodeQL-esq scan results for the file named "J20736", which is vulnerable to CWE-78 (OS command injection):
\begin{lstlisting}[ basicstyle=\scriptsize\ttfamily,
                    columns=fullflexible,
                    keepspaces=true,
                    breaklines=true,
                    frame=single,
                    framerule=0pt,
                    backgroundcolor=\color{gray!10}]
"Uncontrolled command line","Using externally controlled strings in a command line is vulnerable to malicious changes in the strings.","error","This command line depends on a [[""user-provided value""|""relative:///src/testcases/CWE78_OS_Command_Injection/J20736.java:13:65:13:88""]].","/src/testcases/CWE78_OS_Command_Injection/J20736.java","31","53","31","68"
\end{lstlisting}

The SpotBugs-fsb scan results for the file named "J20736", which is vulnerable to CWE-78 (OS command injection):
\begin{lstlisting}[ basicstyle=\scriptsize\ttfamily,
                    columns=fullflexible,
                    keepspaces=true,
                    breaklines=true,
                    frame=single,
                    framerule=0pt,
                    backgroundcolor=\color{gray!10}]
H S SECCI: This usage of java/lang/Runtime.exec(Ljava/lang/String;)Ljava/lang/Process; can be vulnerable to Command Injection  At J20736.java:[line 31]
\end{lstlisting}

The GPT-4 LLM (using $p_{cot-8s}$) scan results for the file named "J20736", which is vulnerable to CWE-78 (OS command injection):
\begin{lstlisting}[ basicstyle=\scriptsize\ttfamily,
                    columns=fullflexible,
                    keepspaces=true,
                    breaklines=true,
                    frame=single,
                    framerule=0pt,
                    backgroundcolor=\color{gray!10}]
1. Identify Weaknesses: The code appears to have a command injection vulnerability.

2. Review User Input Handling: The code takes user input from the HTTP request query string without any validation or sanitization and uses it in a system command.

3. Analyze Data Flow: The untrusted data from the HTTP request query string is tokenized and if a token starts with "id=", the rest of the token is used as data. This data is then appended to a system command and executed.

4. Check for Mitigations: There are no mitigations in place to prevent command injection. The user input is not validated or sanitized before being used in a system command.

5. Evaluate Conditional Branching: There is no dead code in the program. All branches can be reached.

6. Assess Error Handling: The code does not handle any errors that may occur during the execution of the system command.

7. Identify Code Leaking Secrets: The code does not contain any secrets.

8. Provide verdict:
   - vulnerability: YES | vulnerability type: CWE-78 (OS Command Injection) | The code executes a system command with unsanitized user input, leading to a command injection vulnerability.
\end{lstlisting}

The Claude 3 Opus LLM (using $p_{cot-8s}$) scan results for the file named "J20736", which is vulnerable to CWE-78 (OS command injection):
\begin{lstlisting}[ basicstyle=\scriptsize\ttfamily,
                    columns=fullflexible,
                    keepspaces=true,
                    breaklines=true,
                    frame=single,
                    framerule=0pt,
                    backgroundcolor=\color{gray!10}]
Here is my analysis of the provided Java code:

1. Identify Weaknesses: The code may be vulnerable to OS command injection (CWE-78) if the 'data' variable is influenced by user input.

2. Review User Input Handling: The 'data' variable is populated from the query string parameter "id" without any validation or sanitization. This allows user input to directly influence the 'data' variable.

3. Analyze Data Flow: The untrusted 'data' variable is appended to the 'osCommand' string which is then passed to Runtime.exec() to execute a system command. There is no sanitization of 'data' before using it in the command.

4. Check for Mitigations: There are no input validation or sanitization steps to mitigate command injection. The 'data' parameter is used unsafely in the execution of a system command.

5. Evaluate Conditional Branching: The code has a conditional branch based on PRIVATE_STATIC_FINAL_TRUE which is always true. The else block setting data to null is unreachable dead code and not relevant.

6. Assess Error Handling: The code does not contain any explicit error handling for the command execution.

7. Identify Code Leaking Secrets: The code does not appear to leak any secrets.

8. Verdict:
vulnerability: YES | vulnerability type: CWE-78 |

In summary, the code is vulnerable to OS command injection (CWE-78) because the 'data' variable from the "id" query string parameter is used without sanitization to construct a system command that is executed. An attacker could inject arbitrary OS commands via the "id" parameter.
\end{lstlisting}


The CodeQL-esq scan results for the file named "J23877", which is not vulnerable to CWE-89 (SQL injection):
\begin{lstlisting}[ basicstyle=\scriptsize\ttfamily,
                    columns=fullflexible,
                    keepspaces=true,
                    breaklines=true,
                    frame=single,
                    framerule=0pt,
                    backgroundcolor=\color{gray!10}]
"Query built by concatenation with a possibly-untrusted string","Building a SQL or Java Persistence query by concatenating a possibly-untrusted string is vulnerable to insertion of malicious code.","error","Query built by concatenation with [[""this expression""|""relative:///src/testcases/CWE89_SQL_Injection/s02/J23877.java:24:101:24:108""]], which may be untrusted.","/src/testcases/CWE89_SQL_Injection/s02/J23877.java","24","47","24","114"

"Query built by concatenation with a possibly-untrusted string","Building a SQL or Java Persistence query by concatenating a possibly-untrusted string is vulnerable to insertion of malicious code.","error","Query built by concatenation with [[""this expression""|""relative:///src/testcases/CWE89_SQL_Injection/s02/J23877.java:62:101:62:108""]], which may be untrusted.","/src/testcases/CWE89_SQL_Injection/s02/J23877.java","62","47","62","114"
\end{lstlisting}

The SpotBugs-fsb scan results for the file named "J23877", which is not vulnerable to CWE-89 (SQL injection):
\begin{lstlisting}[ basicstyle=\scriptsize\ttfamily,
                    columns=fullflexible,
                    keepspaces=true,
                    breaklines=true,
                    frame=single,
                    framerule=0pt,
                    backgroundcolor=\color{gray!10}]
M S SQL: testcases.CWE89_SQL_Injection.s02.J23877.processG2B(HttpServletRequest, HttpServletResponse) passes a nonconstant String to an execute or addBatch method on an SQL statement  At J23877.java:[line 62]
\end{lstlisting}

The GPT-4 LLM (using $p_{cot-8s}$) scan results for the file named "J23877", which is not vulnerable to CWE-89 (SQL injection):
\begin{lstlisting}[ basicstyle=\scriptsize\ttfamily,
                    columns=fullflexible,
                    keepspaces=true,
                    breaklines=true,
                    frame=single,
                    framerule=0pt,
                    backgroundcolor=\color{gray!10}]
1. Identify Weaknesses: The code has potential SQL Injection vulnerabilities.

2. Review User Input Handling: The user input is taken directly from the HTTP request parameters and used in SQL queries without proper validation or sanitization in the `processB2G` method.

3. Analyze Data Flow: The untrusted data from the HTTP request parameters is directly used in the SQL query in the `processB2G` method.

4. Check for Mitigations: The code uses `PreparedStatement` in the `processB2G` method which mitigates the SQL Injection vulnerability. However, in the `processG2B` method, the code uses `Statement` which is vulnerable to SQL Injection.

5. Evaluate Conditional Branching: There is no dead code in the given code.

6. Assess Error Handling: Errors are logged with a warning level, but the execution continues. This could potentially lead to further errors or unexpected behavior.

7. Identify Code Leaking Secrets: The code does not seem to contain any secrets.

8. Provide verdict:
   - vulnerability: YES | vulnerability type: CWE-89 (SQL Injection) | The `processG2B` method is vulnerable to SQL Injection as it uses `Statement` to execute SQL queries.
   - vulnerability: NO | vulnerability type: CWE-89 (SQL Injection) | The `processB2G` method is not vulnerable to SQL Injection as it uses `PreparedStatement` to execute SQL queries.
\end{lstlisting}

The Claude 3 Opus LLM (using $p_{cot-8s}$) scan results for the file named "J23877", which is not vulnerable to CWE-89 (SQL injection):
\begin{lstlisting}[ basicstyle=\scriptsize\ttfamily,
                    columns=fullflexible,
                    keepspaces=true,
                    breaklines=true,
                    frame=single,
                    framerule=0pt,
                    backgroundcolor=\color{gray!10}]
Here is my analysis of the provided Java code:

1. Identify Weaknesses: The code may be vulnerable to SQL injection in the processG2B method. It directly concatenates user input into SQL queries without proper sanitization.

2. Review User Input Handling: In processG2B, the 'data' variable is hardcoded to "foo" regardless of the conditional branching. It does not come from user input. In processB2G, the 'data' variable comes from the 'name' request parameter, which is user input. However, it is properly parameterized in the SQL query using a PreparedStatement.

3. Analyze Data Flow: In processG2B, the hardcoded 'data' value flows unsanitized into the SQL query. In processB2G, the user input 'data' is safely passed to the SQL query via a parameterized PreparedStatement.

4. Check for Mitigations: The processB2G method mitigates SQL injection by using a parameterized PreparedStatement. The processG2B method has no mitigations and concatenates the 'data' value directly into the query.

5. Evaluate Conditional Branching: The conditional branching in both methods does not affect the data flow or vulnerability. The same code is executed regardless of the branch taken.

6. Assess Error Handling: Database and SQL errors are caught and logged at the WARNING level. The application continues execution after logging.

7. Identify Code Leaking Secrets: No sensitive information like passwords, keys or secrets are leaked in the provided code.

8. Verdict:
vulnerability: YES | vulnerability type: CWE-89 | details: Potential SQL injection in processG2B due to direct concatenation of hardcoded data into SQL query.
vulnerability: NO | vulnerability type: CWE-89 | details: SQL injection mitigated in processB2G by using parameterized PreparedStatement.

In summary, there is a potential SQL injection vulnerability in the processG2B method, but it is not currently exploitable since the concatenated data is hardcoded and not influenced by user input. The processB2G method is not vulnerable to SQL injection due to proper use of parameterized queries.
\end{lstlisting}

\end{document}